\documentclass[aps,prl,twocolumn,amsmath,amssymb,superscriptaddress,nobibnotes,showpacs]{revtex4}

\usepackage{graphicx}
\usepackage{dcolumn}
\usepackage{bm}
\usepackage{amsmath}
\usepackage{amssymb,amsthm}
\usepackage{epsfig,color}
\usepackage[sans]{dsfont}

\definecolor{nblue}{rgb}{0.2,0.2,0.7}
\definecolor{ngreen}{rgb}{0.2,0.6,0.2}
\definecolor{nred}{rgb}{0.7,0.2,0.2}
\definecolor{nblack}{rgb}{0,0,0}

\newcommand{\tr}{\text{tr}}

\def\tr{\mbox{tr}}
\def\bea{\begin{eqnarray}}
\def\eea{\end{eqnarray}}

\begin{document}

\title{No-signaling Principle Can Determine Optimal Quantum State Discrimination}

\author{Joonwoo Bae} \email{bae.joonwoo@gmail.com}
\affiliation{School of Computational Sciences, Korea Institute for Advanced Study, Seoul, 130-012, Republic of Korea,}
\author{Won-Young Hwang}
\affiliation{Department of Physics Education, Chonnam National University, Gwangju 500-757, Republic of Korea,}
\author{Yeong-Deok Han}
\affiliation{Department of Game Contents, Woosuk University, Wanju, Cheonbuk 565-701, Republic of Korea.}

\date{\today}


\begin{abstract}
We provide a general framework of utilizing the no-signaling principle in derivation of the guessing probability in the minimum-error quantum state discrimination. We show that, remarkably, the guessing probability can be determined by the no-signaling principle. This is shown by proving that in the semidefinite programming for the discrimination, the optimality condition corresponds to the constraint that quantum theory cannot be used for a superluminal communication. Finally, a general bound to the guessing probability is presented in a closed form.
\end{abstract}


\maketitle


To find and characterize capabilities of quantum systems in their applications to information processing, often, leads to optimization problems which are in general considered to be difficult. Nevertheless, any optimal performance to be obtained at the end should satisfy fundamental principles that quantum theory fulfills, although it is neither known nor clear if they are tightly related one another, e.g. see Ref. \cite{ref:infcau}. Fundamental principles can be found to be a useful tool to derive \emph{limitations} on optimal quantum performance, in such a way that too good performance would be contradictive.


It turns out that the no-signaling principle, one of the most conservative assumptions in physics, can be used to characterize correlations that are not allowed among parties sharing physical systems \cite{ref:infcau, ref:infcaup}. Assuming quantum systems are shared, consequently, it follows that local operations such as quantum cloning or quantum state discrimination (QSD) cannot work arbitrarily well, since the no-signaling constraint would be violated. This has been considered in specific cases to derive only \emph{bounds} to optimal quantum performances, e.g. \cite{ref:gisin, ref:hwang}. In particular, QSD is of both fundamental and practical importance in wide range of quantum information applications \cite{ref:hel}. It is not only related to fundamental results, such as no-cloning, no-signaling, and non-locality in quantum mechanics \cite{ref:gisin, ref:hwang, ref:masanes}, but also applied to quantum communication or signal processing, e.g. Ref. \cite{ref:hel}.

In this work, we provide a general framework of utilizing the no-signaling principle in derivation of optimal QSD, in such a way that if QSD works better than some threshold (in terms of the guessing probability, or the minimum error), superluminal communication would follow. We show that, remarkably, the obtained threshold actually coincides to that of the optimal QSD. This is shown by proving that the optimality condition in a mathematical formulation for optimal QSD, which will be introduced the semidefinite programming later, corresponds to those in QSD constrained by the no-signaling principle. Hence, the no-signaling constraint turns out to be the physical principle that dictates the optimal performance in discriminating among quantum states. A general and computable bound to optimal QSD is then provided in a closed form. The result also strengthens relations among fundamental no-go theorems in Refs \cite{ref:gisin, ref:hwang, ref:masanes}.

Let us begin by fixing notations. Throughout the paper, $\{q_x, \rho_{x} \}_{x=1}^{N}$ denotes the situation that a quantum state $\rho_{x}$ is generated with \emph{a prior} probability $q_{x}$, where $\sum_{x} q_{x} =1$. Measuring quantum systems is described by Positive-Operator-Valued-Measure (POVM) $\{ M_{x}\}_{x = 1}^{N}$, where i) $M_{x}\geq 0$ for all $x$, and ii) $\sum_{x} M_{x}=I$. Then, the minimum-error QSD among $\{q_x, \rho_{x} \}_{x=1}^{N}$ defines an optimization problem over POVMs such that the error is minimized or equivalently the guessing probability, i.e. the probability of making a correct guess, is maximized. We write $P(x|y)$ the probability that measurement $M_{x}$ is "clicked" when a quantum state $\rho_{y}$ is actually given. The probability measure for quantum states is given by the Born rule, $P(x|y) = \tr[\rho_{y} M_{x}]$, known as Gleason's theorem \cite{ref:gleason}.

\emph{The guessing probability} denotes the maximum probability of correctly guessing, \bea P_{\mathrm{guess}} = \sum_{x=1}^{N} q_{x} P(x|x) = \max_{\{ M_{x}\}_{x =1}^{N}} \sum_{x=1}^{N} q_{x} \tr[\rho_{x}M_{x}]. \label{eq:guess}\eea For the simplest case of two-state discrimination $\{q_{x}, \rho_{x} \}_{x=1}^{N=2}$, the optimal one is known as the Helstrom bound denoted by $P_{\mathrm{guess}}^{( \mathrm{H})}$ as follow, \bea  P_{\mathrm{guess}}^{( \mathrm{H})} = \frac{1}{2} (1 + \|q_1 \rho_1 - q_2 \rho_2  \|). \label{eq:helstrom} \eea For more than two quantum states, the guessing probability is known only in restricted cases, for instance, geometrically uniform states \cite{ref:eldar}.

We now approach to the QSD problem with a fundamental constraint, the no-signaling principle, that should be fulfilled in any information processing by quantum systems. The main idea is to incorporate the QSD problem to a communication scenario between two parties, Alice and Bob who attempt to communicate by making only use of shared quantum states and measurement. Then, the optimal performance of local operations and local measurements can be limited by the no-signaling constraint.

Let us consider the following communication protocol, where two parties share copies of a quantum state $|\psi\rangle_{AB}$. Suppose that Alice encodes a message $x\in \{1,\cdots,N\}$ into the application of one of POVMs, $M^{(A,x)} =\{ M_{y}^{(A,x)},~ y=1,\cdots \}$, in which $M^{(A,x)}$ is complete, i.e., $ \sum_{y} M_{y}^{(A,x)} =I$ for each $x=1,\cdots,N$. For instance, the message $x$ is encoded in the application of the complete POVM $M^{(A,x)}$. The resulting state in the Bob's side is one of those states, $\rho_{y}^{(x)} =(p_{y}^{(x)})^{-1} \tr_{A} | \psi \rangle_{AB}\langle \psi| (M_{y}^{(A,x)} \otimes I)$ with probability $p_{y}^{(x)} =  \langle\psi |  (M_{y}^{(A,x)} \otimes I) |\psi\rangle$. Given that the measurement outcome is not announced, Bob only knows his system is in $\rho_{y}^{(x)}$ with probability $p_{y}^{(x)}$, that is, described by a mixed state, $\rho_{B}^{(x)} =\sum_{y} p_{y}^{(x)} \rho_{y}^{(x)}$. For another message $x'$ corresponding to the application of $M^{(A, x')}$, Bob's system results in the state $\rho_{B}^{(x')}$, which is equal to $\rho_{B}^{(x)}$ while they are in different mixtures. In fact, using an appropriate POVM, Alice can prepare any quantum states on Bob's side, i.e. any state-decomposition in the Bob's ensemble. This is known as the Gisin-Hughston-Jozsa-Wootters (GHJW) theorem \cite{ref:ghjw}. Since they are identical quantum states, Bob can never learn about the POVM Alice has applied, and consequently, no message is allowed to be transferred in this way.

Now, let $P_{D}(x|x')$ denote the probability that Bob's detector gives an answer $x$ when Alice has applied measurement $M^{(A,x')}$. Note the normalization condition that for each $x'$, it holds that $\sum_{x}P_{D}(x|x') = 1$. We also note that Bob's device for the discrimination is not specified, but only its input-output relation. This can be thought of as a black box scenario.

It is clear that if the no-signaling constraint is to be fulfilled, the input-output relation cannot be given arbitrarily. Suppose that $\sum_{x} P_{D} (x|x) >1$, meaning that $\sum_{x} [P_{D} (x|x) - P_{D} (x|x')]>0$ for some $x'\neq x$, from which there would be at least a single $x$ such that $P_{D} (x|x) > P_{D} (x|x')$. This immediately implies that $P_{D}$ is not non-signaling \cite{ref:masanes} since superluminal communication can be constructed in the following way: if Alice applies two POVMs $M^{(A,x)}$ and $M^{(A,x')}$ to encode $0$ and $1$, respectively, Bob finds from his detector how frequently the outcome $x$ appears and then conclude if Alice's encoding is $0$ or $1$. Therefore, from the no-signaling constraint, we have \bea \sum_{x} P_{D} (x|x ) \leq 1, \label{eq:nosig} \eea on the Bob's detector for the discrimination.

We now relate QSD of $\{q_x,\rho_x \}_{x=1}^{N}$ to the communication in the above. The key idea is to consider the case that, one of POVM elements in the set $M^{(A,x)}$, say the first element $M_{1}^{(A,x)}$ prepares state $\rho_x$ on the Bob's side with probability $p_{x}$, and the rest $I-\sum_{y\neq1} M_{y}^{(A,x)}$ does a state $\sigma_{x}$, \bea \rho_{B}^{(x)} = p_{x} \rho_{x} + (1-p_{x}) \sigma_{x},~~~\mathrm{for} ~ x=1,\cdots,N. \label{eq:bobi} \eea  Remind that this is always possible from the GHJW theorem \cite{ref:ghjw}. If QSD among $\{\rho_x \}_{x=1}^{N}$ works too well, Eq. (\ref{eq:nosig}) would be violated, meaning that the no-signaling constraint is not fulfilled. In this way, the no-signaling principle can constrain the guessing probability for QSD among $\{\rho_x \}_{x=1}^{N}$.

In what follows, we derive a threshold of the guessing probability in QSD among $\{\rho_x \}_{x=1}^{N}$ in such a way that the no-signaling principle is not violated. From Eq. (\ref{eq:bobi}), it follows that $p_x P(x|x) \leq P_{D} (x|x)$ since, for Alice's measurement $M^{(A,x)}$, the probability that Bob's detector answers $x$ consists of contributions both by the state $\rho_x$ with probability $p_x$ and the rest from state $\sigma_x$. Then, the no-signaling constraint in Eq. (\ref{eq:nosig}) leads to the following bound, \bea \sum_{x} p_{x} P(x|x) \leq 1.  \label{eq:nosig2}\eea Recall that Bob's measurement device for the discrimination is not specified, but only its input-output relation - like a black box scenario. Note also that the probability measure, Born rule is not applied yet. The bound in Eq. (\ref{eq:nosig2}) is only the condition that the guessing probabilities $P(x|x)$ do not lead to superluminal communication. The followings are assumed, so far: \emph{a) bipartite quantum states, b) the Born rule to the Alice's system}, and \emph{c) the no-signaling principle between the two parties}.

In fact, in the above scenario, the bound obtained in Eq. (\ref{eq:nosig2}) corresponds to QSD among $\{q_x, \rho_{x}\}_{x=1}^{N}$ where, \bea q_{x} = \frac{p_{x}}{\sum_{x'=1}^{N} p_{x'}} \label{eq:qp}. \eea This is because, Bob's device aims at discriminating among $\{\rho_{x}\}_{x=1}^{N}$, and the \emph{a priori} probability that state $\rho_{x}$ appears from $\{\rho_{x}\}_{x=1}^{N}$ can be found as $q_x$ in the above. Having collected all these, it is straightforward to derive the main result. \\


\textbf{Proposition.} From the no-signaling principle, the guessing probability in QSD among $\{q_{x},\rho_{x}\}_{x=1}^{N}$ must be bounded as follows, \bea P_{\mathrm{guess}} = \sum_{x} q_{x} P(x|x) \leq \frac{1}{\sum_{x} p_{x}},\label{eq:result}\eea where $\{p_{x}\}_{x=1}^{N}$ are from the identical ensembles in Eq. (\ref{eq:bobi}) with the relation in Eq. (\ref{eq:qp}). The equality holds when the equality in Eq. (\ref{eq:nosig2}) hold for all $x=1,\cdots,N$\\

The equality in Eq. (\ref{eq:nosig2}) means that, Bob's measurement device works in a way that for each ensemble $\rho_{B}^{(x)}$ (see Eq. (\ref{eq:bobi})), the measurement device responds only to $\{\rho_{x}\}_{x=1}^{N}$ but not  $\{\sigma_{x} \}_{x=1}^{N}$. Therefore, the condition that the equality in Eq. (\ref{eq:result}) holds is the existence of identical ensembles in Eq. (\ref{eq:bobi}) such that the measurement device only responds to those states $\{\rho_{x}\}_{x=1}^{N}$ but not $\{\sigma_{x}\}_{x=1}^{N}$. Taking the measurement postulate in quantum theory into account (see Eq. (\ref{eq:guess})), the condition of the equality in Eq. (\ref{eq:result}) means the existence of POVM $\{M_{x}\}_{x=1}^{N}$ and $\{\sigma_{x}\}_{x=1}^{N}$ such that, for all $x = 1,\cdots,N$,
\bea \sum_{x} p_{x}\tr[\rho_{x} M_{x}] = 1,~\mathrm{or}~ \mathrm{equivalently},~  \tr[\sigma_{x} M_{x}] = 0. \label{eq:opts} \eea When each state $\sigma_{x}$ satisfies the condition in the above with respect to POVM $\{M_{x}\}_{x=1}^{N}$, we call it \emph{complementary} to $\rho_{x}$. This defines the relation between $\rho_{x}$ and $\sigma_{x}$ in the ensemble in Eq. (\ref{eq:bobi}) for the inequality in Eq. (\ref{eq:result}) to be saturated.

To summarize what we have shown so far, a general framework for utilizing the no-signaling principle in QSD among $\{q_x, \rho_x \}_{x=1}^{N}$ is presented, and a general bound is also obtained in Eq. (\ref{eq:result}). The equality also holds if \emph{complementary states $\{\sigma_{x}\}_{x=1}^{N}$ exist for given states $\{\rho_{x}\}_{x=1}^{N}$ to be discriminated among, i.e., i) the measurement device does not respond to these states, (see Eq. (\ref{eq:opts}) under the assumption of the Born rule) ii) the identical ensembles in Eq. (\ref{eq:bobi}) can be found fulfilling the relation between $p_{x}$ and $q_{x}$ in Eq. (\ref{eq:qp}).} Once the equality holds, it is also crucial to know if \emph{iii) the bound coincides to the guessing probability of optimal QSD.}

In the rest of the paper, we answer to three questions addressed in the above. Namely, we show that for any optimal QSD, one can find identical ensembles in Eq. (\ref{eq:bobi}) fulfilling i), ii), and iii). This leads to the following conclusion.\\

\emph{The guessing probability of optimal QSD can be determined by the no-signaling principle.}\\

To proceed the proof, we consider the optimality condition of the semidefinite programming (SDP) for the guessing probability of optimal QSD. In an SDP, an optimization problem can be written in two forms, called primal and dual, and each one is called feasible when variables satisfying given constraints are not of an empty set \cite{ref:sdpbook}. When both problems are feasible, it follows that optimal solutions exist and can be obtained by solving either form of the problem.

There are so-called Karush-Kuhn-Tucker (KKT) conditions which can also decide if an optimal solution exists in an SDP problem. In fact, variables satisfying the KKT conditions give an optimal solution of both primal and dual problems. In summary, optimal solutions can be obtained in either way: i) solving KKT conditions or ii) solving either primal or dual problem in which both are feasible. The KKT conditions often form non-linear, and are therefore considered not to be easier to solve than to do a primal or dual problem.

We now show that the lists i), ii), and iii) correspond to the optimality condition, i.e. the KKT, of the SDP for the guessing probability in optimal QSD. \\

\emph{Proof of the result.} Let us start by formulating the SDP for the guessing probability of optimal QSD among $\{q_x, \rho_x \}_{x=1}^{N}$ as follows, what we call the primal problem,
\bea \max && f(\{M_x \}_{x=1}^{N}) = \sum_{x} q_x \tr[ \rho_x M_x]  \nonumber  \\
\mathrm{subject}~\mathrm{to} &&  M_x \geq 0, ~~ \sum_x M_x = I, \label{eq:pri}  \eea where POVM $\{ M_{x}\}_{x=1}^{N}$ are called primal variables. The Lagrangian can be constructed as \bea &&L(\{M_x\}_{x=1}^{N}, \{\sigma_x\}_{x=1}^{N}, K) = \label{eq:lag} \\ && f(\{M_x\}_{x=1}^{N}) - \sum_x \tr[\sigma_x M_x] + \tr[K(\sum_x M_x -I)], \nonumber \eea with non-negative operators $\{\sigma_{x}\}_{x=1}^{N}$ and $K$ called dual variables. It is also straightforward to derive the dual problem \cite{ref:sdpbook},
\bea \min && \tr[K] \nonumber  \\
\mathrm{subject}~\mathrm{to} && K\geq q_x \rho_x, ~~\forall~ x=1,\cdots,N. \label{eq:dual}  \eea It is clear that primal and dual problems are feasible, and therefore optimal solutions exist and can be found by solving either form of the problem.

The optimal solutions can also be obtained by solving the KKT conditions, which are obtained from the Lagrangian in Eq. (\ref{eq:lag}):
\bea \tr[\sigma_{x}M_{x}] & = & 0, ~ \mathrm{and}  \label{eq:consig} \\
K  & = & q_x \rho_x + \sigma_x,~~ \forall ~ x=1,\cdots,N,~ \label{eq:eqesb} \eea and the constraints in Eqs. (\ref{eq:pri}) and (\ref{eq:dual}). Note that existence of optimal solutions is already guaranteed by the fact that both the primal and the dual problems are feasible.

We are now ready to show that, when the equality in Eq. (\ref{eq:result}) is saturated, the guessing probability corresponds to that of optimal QSD. First, the condition in Eq. (\ref{eq:consig}) called \emph{complementary slackness} means that each optimal $M_{x}$ is orthogonal to dual variable $\sigma_{x}$. Existence of states $\{\sigma_{x}\}_{x=1}^{N}$ that satisfy the condition in Eq. (\ref{eq:opts}) is therefore shown. Second, the condition in Eq. (\ref{eq:eqesb}) assures the existence of an identical ensemble that can be decomposed $N$ different ways such that each decomposition consists of one of states $\{\rho_x\}_{x=1}^{N}$ and its corresponding complementary state $\sigma_x$, as it is shown in Eq. (\ref{eq:bobi}). After normalization $\widetilde{K} = K/\tr[K]$, the identical ensembles $\widetilde{K}$ can be explicitly seen,
\bea
\widetilde{K} = \frac{q_x}{\tr[K]} \rho_x + \frac{1}{\tr[K]}\sigma_x, ~~\forall~ x=1,\cdots,N. \label{eq:tilK}
\eea Hence, the existence of an identical ensemble in Eq. (\ref{eq:bobi}) together with complementary states $\{\sigma_{x}\}_{x=1}^{N}$ is shown. Finally, remind that the solution of the dual problem in Eq.(\ref{eq:dual}) is given by $\tr[K]$. The ensemble in Eq. (\ref{eq:tilK}) has the state $\rho_{x}$ with probability $q_{x}/\tr[K]$, which corresponds to $p_{x}$ in Eq. (\ref{eq:bobi}). From the normalization $\sum_{x}q_{x} =1$, it follows that $ \tr[K] = 1/\sum_{x} p_{x}$, which coincides to the upper bound in Eq. (\ref{eq:result}) obtained by the no-signaling constraint. Therefore, the bound in Eq. (\ref{eq:result}) is shown to be indeed the guessing probability in optimal QSD. $\Box$\\

A general bound to the guessing probability can be derived using the condition of the identical ensemble in Eq. (\ref{eq:bobi}): for all $x,y$, \bea \| p_x \rho_x - p_y \rho_y \| = \|(1-p_x)\sigma_x - (1-p_y)\sigma_y \|. \label{eq:final} \eea From this, one can compute the quantity, $\sum_{x}p_{x}$, in Eq. (\ref{eq:result}). Here, we derive a very general bound from the fact that in Eq. (\ref{eq:final}) r.h.s. is not larger than $2 - (p_{x} + p_{y})$, and l.h.s. is equal to, $(\sum_{z} p_{z})\|q_{x}\rho_{x} - q_{y}\rho_{y}\|$. As a result, we have, \bea P_{\mathrm{guess}} \geq \frac{1}{N} (1 + \frac{1}{2}\sum_{x=1}^{N} \| q_{x}\rho_{x} - q_{x+1}\rho_{x+1} \| ), \nonumber \eea where $p_{N+1}=p_{1}$ and $\rho_{N+1}=\rho_{1}$. Although this bound is in general not tight, in particular when $N$ exceeds to the dimension of the Hilbert space supporting quantum states $\{ \rho_{x}\}_{x=1}^{N}$, the usefulness of this bound is especially worthy of notice as no assumption is made on both the structure among given quantum states and the \emph{a priori} probabilities. For two-state discrimination, this bound actually coincides to the optimal one, Helstrom bound in Eq. (\ref{eq:helstrom}).


To summarize, we have provided a general framework of utilizing the no-signaling principle in QSD problems. It is shown that the guessing probability in optimal QSD can be determined by the no-signaling principle, i.e. \emph{the no-signaling constraint is the physical principle that dictates the optimal performance in QSD}. We also highlight the methodology employed, that the no-signaling principle is related to the optimality condition (i.e. KKT) of the SDP problem for QSD. This may envisage a usefulness of the SDP in quantum optimization problems as a method of characterizing physical principles that dictate optimal quantum performances. In this way, the guessing probability is obtained without resort to the measurement postulate via the Born rule, as follows.

Recall the list, a), b), and c), assumed when deriving the guessing probability (with the equality) in Eq. (\ref{eq:result}). Note that the measurement postulate on Bob's quantum states is not assumed. Probabilities saturating the equality in Eq. (\ref{eq:result}) are actually obtained by imposing the no-signaling constraint to Bob's probabilities. Then, from SDP it is shown that there always exist POVMs that attain Bob's probabilities from his states via the Born rule. This shows that Gleason's theorem for any set of quantum states $\{ q_{x},\rho_x\}_{x=1}^{N}$ is derived from the three assumptions. This is in fact the converse of the recent result on the bipartite Gleason correlations \cite{ref:acin1, ref:acin}: any non-signaling correlations between two systems for which local quantum measurements are possible can also be obtained by measurement on some bipartite quantum states. It would be interesting to derive a general proof of the converse: by assuming bipartite quantum states, local quantum measurement on Alice, and the no-signaling constraint between two parties, can Gleason's theorem for Bob's local quantum mechanics be derived?

The guessing probability is connected to the min-entropy, through which the max-entropy quantifying the so-called decoupling approach is also related \cite{ref:renner}. Recently, the connection of the guessing probability to quantum non-locality is shown via the no-signaling principle \cite{ref:jon}. It would be interesting to investigate further operational relations between these entropic quantities and fundamental principles in physics.

This work is supported by the National Research Foundation of Korea (2010-0007208 and KRF-2008-313-C00185) and Woosuk University. J.B. also thanks the Institut Mittag-Leffler (Djursholm, Sweden) for the support during his visit.


\end{document}